\documentclass[useAMS,usenatbib]{mn2e}

\usepackage{graphicx}

\title[Wide double stars - I. The spectroscopic binaries]{Double stars with wide separations in the AGK3 - I. The 
components that are themselves spectroscopic binaries\footnotemark[1]\thanks{based on observations performed
at the Observatoire de Haute--Provence (CNRS), France}}
\author[J.-L. Halbwachs, M. Mayor and S. Udry]{J.-L. Halbwachs$^{1}$\thanks{E-mail:
jean-louis.halbwachs@astro.unistra.fr}, M. Mayor$^{2}$ and S. 
Udry$^{2}$\\
$^{1}$Observatoire Astronomique de Strasbourg (UMR 7550),
   11 rue de l'Universit\'{e}, F--67\,000 Strasbourg, France\\
$^{2}$Observatoire Astronomique de l'Universit\'e de Gen\`eve, 51, chemin des maillettes,
CH--1290 Sauverny, Switzerland}
\begin{document}

\date{Accepted . Received 2011 ; in original form 2011}

\pagerange{\pageref{firstpage}--\pageref{lastpage}} \pubyear{2011}

\maketitle

\label{firstpage}

\begin{abstract}
Wide binaries are tracers of the gravity field of the Galaxy, but their study requires some caution.
A large list of common proper motion stars selected from the AGK3 were monitored with the {\sc coravel}
spectrovelocimeter, in order to prepare a sample of physical binaries with very wide separations.
Sixty-six stars received
special attention, since their RV seemed to be variable. These stars were monitored over
several years in order to derive the elements of their spectroscopic orbits. In addition, 10 of them received accurate
RV measurements from the {\sc sophie} spectrograph of the T193 telescope at the Observatory
of Haute-Provence.

For deriving the orbital elements of double-lined spectroscopic binaries (SB2), a new method was applied,
which assumed that the RV of blended measurements are linear combinations of the RV of
the components. Thirteen SB2 orbits were thus calculated.

The orbital elements were eventually obtained for 52 spectroscopic binaries (SB), two of them making a triple system.
Forty SB
received their first orbit, and the orbital
elements were improved for 10 others. In addition, 11 spectroscopic binaries were discovered with very long periods
for which the orbital parameters were not found. The median period of the 40 first orbits is 1 year,
and several SB should be resolved or should receive an astrometric orbit in the future,
providing the masses of the components. In addition, it appeared that HD 153252 has a close
companion which is a candidate brown dwarf with a minimum mass of 50 Jupiter masses.

The final selection of the wide binaries and the derivation of their statistical properties
will be presented in a second paper.
\end{abstract}


\section{Introduction}


Binarity is a very common stellar property, which covers a very wide range of separations,
from two stellar radii to thousands of astronomical units. Close binaries have short
orbital periods (from a few hours to a few years), and they are easily detected from variations
in their radial velocities (RV). For that reason, they are the most commonly studied of this class
of stars (see, eg, \citealt{Halbwachs03}). 
Binaries with periods around a few centuries are also rather well
known: the components are sufficiently separated to be detected visually (in the past)
or on images, and they are still close enough to avoid the risk of confusing a field star
and a binary component: they may be discarded by applying a statistical criterion almost as
old as the discovery of double stars \citep{Struve52}. 

Very wide binaries are particularly interesting, since the distribution of their
separations is a clue to the gravitational perturbers which made the gravity field of the Galaxy
(see \citealt{Jiang10} and references therein). However, the
selection of binaries with separations of thousands of astronomical units
is difficult: the components have also wide apparent separations, and it is necessary
to use additional criteria in order to discard the optical companions. Trigonometric
parallax, or another estimation of the distance, is sometimes used, but the most efficient,
and the most employed criterion is proper motion: when the semi-major axis of the orbit is large,
the orbital motion of the stars around the barycentre of the system generates a difference
in proper motion which is negligible, and the components must have similar apparent displacements.
Such binaries are called common proper motion (CPM) stars.

In the past, this criterion was used by Luyten to search for wide systems by visual inspection
of photographic plates; his prospect was performed over more than 40 years, from \citet{Luyten40}
to \citet{Luyten87} and led to the discovery of 6121 systems. However, despite their large number,
these systems were not used to derive the properties of wide binaries; the main reason is that
the selection was based on subjective criteria, and, as a consequence, it would be hard to
estimate its incompleteness, and also the contamination by optical pairs. Moreover, since the
components of Luyten's double stars are rather faint (around 16th magnitude), it would be difficult to
improve the selection with complementary data.

Another selection of CPM pairs was performed by \citet{Halbwachs86}, on the basis of the
AGK2/3 catalogue \citep{Lacroute74}. Four hundred and thirty-nine CPM double stars were selected, 
and the number of optical
pairs was estimated to be around 40. Since the AGK3 stars are brighter than 12th magnitude, 
it was decided to select the physical binaries on the basis of the RV of the components.
As is the case for the proper motions, the RV are marginally affected by the orbital motion, and the
components of wide binaries are therefore also common RV stars. Two hundred and sixty-six stars were
selected, and they were measured with the spectrovelocimeter {\sc coravel}. 
This programme was initiated 
in 1986, and it was initially supposed to end after a few years. However, several stars had
variable velocities, since these wide pair components were themselves spectroscopic binaries (SB).
It was then necessary to extend the observations over about 20 years, in order to derive the SB orbital elements, and
thus the systemic velocity of these stars. The present paper is devoted to these 
variable velocity stars, and to the derivation of their SB elements. The selection of the
wide binaries and the derivation of their statistical properties will be treated in a
second paper.

The paper is organized as follows: the RV observations are presented in 
section~\ref{sec:RV}, and the catalogue of the RV measurements is in section~\ref{sec:RVcat}. The calculation of the SB orbits is in section~\ref{sec:SB}; for the double--lined SB (SB2), we present a method to take into account the measurements related 
to blends of the two components. Some interesting points relating to the new SB that we have discovered are discussed in the conclusion, 
section~\ref{sec:conclusion}.
 

\section{The radial velocity measurements}
\label{sec:RV}

\subsection{The {\sc coravel} observations}

The observational programme concerned a large part of the CPM stars listed in \citet{Halbwachs86}.
In that paper, the CPM stars were presented in two tables, according to the time $T=\theta/\mu$,
where $\theta$ is the apparent separation and $\mu$ is the proper motion of the pair of stars.
Three hundred and twenty-six pairs of CPM stars with $T < 1000$~years are listed in the first table,
and the estimated frequency of optical pairs is only 1.3~\%. The second table contains 113 CPM
pairs with $T$ between 1000 and 3500 years; this range corresponds to an expected frequency of optical pairs 
of 40~\%. Since the proportion of optical pairs is rather important in the second
table, all these 113 CPM pairs were selected in a first step. For the first table, it was
decided to observe only the components of pairs with a primary brighter than 9.15 mag.
This limit was chosen since it appeared that the AGK 2/3 is not complete beyond that magnitude. 

The spectrovelocimeter {\sc coravel} \citep{Baranne79} is installed on the Swiss 1-m telescope,
at the Observatory of Haute-Provence (OHP). For slow rotators with spectral types around K0,
it may provide RV measurements with a precision around 0.3~km/s. However, it
becomes inefficient for stars earlier than F5, except when their spectra contain metallic lines.
Therefore, it is possible to get RV with a precision near 1~km/s for the Am--type stars. Due to 
this restriction, our final programme contains 266 stars observable with {\sc coravel}: 90 are extracted from
the first table, and 176 from the second one.

The programme started in 1986, but several stars were already measured with {\sc coravel} at this
time. The observations of the stars with a constant velocity were stopped after 1 or 2 years, but
several SB were detected, and it was necessary to prolong the programme in order to cover
their periods and to derive their orbital elements. In practice, when a SB was observed, its wide companion was usually observed too. The {\sc coravel} observations were performed until the decommissioning of
that instrument in 2000. 
They led to the detection of 66 stars with variable radial velocity, according to the
$P(\chi^2)$ test at the 1~\% threshold.

\subsection{The {\sc sophie} measurements}

In 2007, we had the opportunity to obtain very accurate RV measurements with the {\sc sophie}
spectrograph installed on the 1.93~m telescope at OHP \citep{Perruchot08, Bouchy09}.
Ten stars with variable RV, but for which
it was still not possible to derive an orbit, were observed in {\it service mode} during 2
semesters. The spectra were registered in high--resolution mode with a signal-to-noise ratio between 20 and 200, depending on the brightness of the star. 
Thanks to the {\sc sophie} automatic pipeline,
they were cross--correlated with a mask close to the actual spectral type of the star. 
For the single-lined binaries (SB1), the RV was directly obtained from the pipeline. For
two SB2, the cross--correlation function (CCF) was fitted with two normal
distributions in order to derive the velocities of both components.

The automatic pipeline provides also RV uncertainties, but they are obviously underestimated.
\citet{Boisse10} estimate that 3 terms must be quadratically added: the instrumental drift is
around 3~m~s$^{-1}$, the guiding error is around 4~m~s$^{-1}$, and they derived from their own observations
an additional error of 8~m~s$^{-1}$. We finally obtain an additional error of 9.4~m~s$^{-1}$, that is
rounded up to 10~m~s$^{-1}$. It is worth noticing that this error does not
apply to the radial velocity with respect to the Sun, but to variations of RV
due to an exoplanetary companion, ie to RV fluctuations over a range 
not larger than around 100~m~s$^{-1}$. Since we are observing SB with stellar components, the
semi-amplitude of the RV is as large as several km~s$^{-1}$, and the error could be larger
than 10~m~s$^{-1}$.
Moreover, we must introduce an offset between the {\sc coravel} measurements and the 
{\sc sophie} measurements. This offset depends on the star, and on the mask used in the
{\sc sophie} reduction. It may be computed with the orbital elements when several {\sc sophie}
measurements are available, but, for a single measurement, it results in an error of
0.3~km~s$^{-1}$, to a rough estimation.

\section{The RV catalogue}
\label{sec:RVcat}

In addition to the observations performed with {\sc coravel} and with {\sc sophie}, we still obtained 3
measurements from two other telescopes: the Euler telescope with the {\sc coralie} spectrovelocimeter, in
La Silla, and the 1~m telescope of the Simeis Observatory in Crimea. Finally, 2275 measurements were
obtained for the 66 stars that were selected as variable or probably variable.
The measurements are gathered in one plain text file, with one header record preceding the RV of each star.
Each header contains:
\begin{itemize}
\item
{\it CPM}, the identification of the star in \citet{Halbwachs86}, which consists in the number of the list, followed by the
number of the pair, followed by ``A'' or ``B'' for the component.
\item
{\it AG}, the AG identification.
\item
$B-V$, the color index assumed in the derivation of the {\sc coravel} radial velocities.
\item
$N_{rec}$, the number of measurements following the header.
\item
$\bar{V}$, the mean radial velocity of the star. When a spectroscopic orbit was derived, as indicated
by a ``O'' in the variability status, $\bar{V}$ is the velocity of the system. Otherwise, it is
an average of the {\sc coravel} measurements with a zero or a blank in the ``c'' column of table~\ref{tab:mesRV},
as explained in section~\ref{sec:SBnoOrb}.
\item
$\sigma_{\bar{V}}$, the uncertainty of $\bar{V}$.
\item
$P(\chi^2)$, the probability to get a $\chi^2$ larger than the one actually obtained, assuming 
the RV is constant in reality. The $\chi^2$ was computed from the  {\sc coravel}
measurements of the primary component.
\item
The final variability status of the star. The following status were found: ``CST?'' for a star which 
could have a constant RV, ``VAR'' when the RV seems to be variable, ``SB1'' or ``SB2'' when at least a
part of the velocity curve is visible, ``SB1O'' and ``SB2O'' when the orbital elements were derived.
\item
A ``$+$'' follows the variability status SB1O or SB2O when RV measurements from an external source
were taken into account to derive the orbital elements. These additional measurements were found in the
SB9 on-line catalogue \citep{Pourbaix04}; they are not 
reproduced here, but the references are given in the notes, section~\ref{sec:notes}.
\item
The right ascension of the ``A'' component, in hours and minutes, is given in order to facilitate the
search of the stars.
\end{itemize}

Some headers are presented in Table~\ref{tab:header}.

\begin{table*}
 \centering
 \begin{minipage}{140mm}
  \caption{Sample of the RV catalogue. The headers of the first six stars; in the catalogue file, each
header is immediately followed by $N_{meas}$ measurements.}
  \begin{tabular}{@{}lcccrcclc@{}}
  \hline
CPM &   AG & $B-V$ & $N_{meas}$ &  $\bar{V}$ & $\sigma_{\bar{V}}$ & $P(\chi^2)$ & Var & $\alpha$ \\
  \hline
2 $\;\;$5A &+46  $\;$87 & 0.58 & 33 &  -5.484 & 0.142 & 0.000 &SB2O & 00 37\\
2 $\;\;$5B &+46  $\;$86 & 1.93 & 19 & -15.714 & 0.370 & 0.000 &VAR & 00 37\\
2 $\;\;$7A &+31  $\;$55 & 0.68 & 43 & -38.802 & 0.142 & 0.000 &SB1O & 00 41\\
2 $\;\;$8A &$-$01  $\;$87 & 0.53 & 51 &   8.328 & 0.055 & 0.000 &SB1O & 00 58\\
2 $\;\;$8B &$-$01  $\;$90 & 0.68 & 14 &  10.431 & 0.195 & 0.001 &SB1  & 00 58\\
1  18A & +31  132 & 0.68 & 90 &  17.204 & 0.075 & 0.000 & SB2O+ & 01 22 \\
  \hline
\label{tab:header}
\end{tabular}
\end{minipage}
\end{table*}

The headers are followed with the measurement records (Table~\ref{tab:mesRV}), which consist of:

\begin{itemize}
\item
The epoch of the observation, in barycentric Julian days, counted since JD~$2\,400\,000$.
The decimal part is restricted to 3 digits for the {\sc coravel} measurements, but 4
digits for the {\sc sophie} measurements.
\item
$V_R$, the RV in~km~s$^{-1}$. The last digit corresponds to 10~m~s$^{-1}$
for the {\sc coravel} measurements, but to 1~m~s$^{-1}$ for the {\sc sophie} measurements.
As often as possible, the {\sc sophie} measurements were corrected for a systematic shift 
between them and the {\sc coravel} ones; the correction is then indicated in 
the notes, section~\ref{sec:notes}.
\item
$\sigma_{RV}$, the uncertainty of $V_R$, in~km~s$^{-1}$, with the same number of digits as $V_R$.
\item
c, an index indicating the measured component of the SB2: ``1'' for the primary, ``2'' for the
secondary, and ``0'' for a blend of both components; blank for SB1.
\item
f, a flag coded as follows: ``R'' when the measurement was rejected in the final calculation, and
``F'' when it applies to a component with a fixed velocity. 
\item
T, a flag indicating the telescope used for the measurement: ``C''= 1~m/{\sc coravel} telescope in Haute-Provence (2208 occurrences),
``E''= {\sc Euler/coralie} telescope at La Silla, Chile (2 occurrences), ``S''= T193/{\sc sophie} (62 occurrences) and ``T''=
1~m telescope of the Simeis Observatory, in Crimea (1 measurement).
\end{itemize}

\begin{table}
 \centering
  \caption{Sample of the RV catalogue. Some measurements following the headers of two of the stars, a
SB1 and a SB2.}
  \begin{tabular}{@{}lrlccc@{}}
JD   & \multicolumn{1}{c}{$V_R$}  &  $\sigma_{RV}$ & c & f & T \\
+2\,400\,000 & km~s$^{-1}$ & km~s$^{-1}$ & & & \\
\hline
48053.561   &-16.57 $\;$ &0.31 $\;$ & & R& C \\
48139.335   &-17.65 $\;$ &0.36 $\;$ & &  & C \\
48878.544   &-18.62 $\;$ &0.30 $\;$ & &  & C \\
49083.925   &-17.77 $\;$ &0.32 $\;$ & &  & C \\
54214.6265  &-17.568 &0.010 & &  & S \\
54255.5037  &-17.551 &0.010 & &  & S \\
\hline
50548.336  & -10.58 $\;$ &0.41 $\;$& 0 & & C \\
50819.550  & -12.44 $\;$ &0.35 $\;$& 0 & & C \\
51197.532  & -17.86 $\;$ &0.53 $\;$& 1 & & C \\
51197.532  &  -4.52 $\;$ &1.07 $\;$& 2 & & C \\
51519.512  & -15.35 $\;$ &0.32 $\;$& 1 & & T \\
51616.605  &   8.56 $\;$ &0.05 $\;$& 1 & & E \\
\hline
\label{tab:mesRV}
\end{tabular}
\end{table}

\begin{table*}
 \centering
 \begin{minipage}{178mm}
  \caption{The average velocities of the stars classified as variable on the basis of $P(\chi^2)<1$~\%.
$N_{CORA}$ is the number of {\sc coravel} observations, and $\Delta T_{CORA}$
is the timespan of these observations; the three columns that follow are statistical
quantities referring to these $N_{CORA}$ measurements of the primary component or of blends: 
$I$ is their ``internal'' error, ie an estimation of
the average $\sigma_{RV}$ derived from the mean of the weights, $E/I$ is the external to internal errors
ratio, where $E$ is the standard deviation of $V_R$, and $P(\chi^2)$ is the probability to get a so large
value of $E/I$ when the RV of the star is constant in reality. $\Delta T_{meas}$ is the timespan of all
the observations, when other sources are added; $N_{meas}$ is the total number of RV measurements (all
sources and all components); for readability, $N_{meas}$ and $\Delta T_{meas}$ are given only when different
from $N_{CORA}$ or $\Delta T_{CORA}$, respectively. An asterisk indicates when the SB orbit is the first
for this object, and $\bar{V}$ is the velocity of the system, in km s$^{-1}$.
}
\scriptsize
  \begin{tabular}{@{}lccrrrrrrrlcr@{}}
  \hline
CPM & HD/BD/HIP & RA & $N_{CORA}$ & $\Delta T_{CORA}$ & $I$ & $E/I$ & $P(\chi^2)$ & $N_{meas}$ &$\Delta T_{meas}$ & Var & $1^{st}$ orbit & $\bar{V}$  \\
 \hline
2:  5A&BD +45 172 & 00 37&  21 &4160 & 0.456&   7.40 &0.000 &  33 & 7481&SB2O &*&$  -5.484\pm  0.142 $\\
2:  5B&BD +45 171 & 00 37&  19 &4160 & 0.606&   2.66 &0.000 &     &     &VAR  & &$ -15.714\pm  1.568 $\\
2:  7A&HD 4153 & 00 41&  43 &4396 & 0.350&  32.92 &0.000 &     &     &SB1O &*&$ -38.802\pm  0.142 $\\
2:  8A&HD 5947 & 00 58&  51 &4461 & 0.445&  17.38 &0.000 &     &     &SB1O &*&$   8.328\pm  0.055 $\\
2:  8B&BD -01 133& 00 58&  11 &4316 & 0.379&   1.70 &0.001 &  14 & 7729&SB1  & &$  10.431\pm  0.615 $\\
1: 18A&HD 8624 & 01 22&  47 &3291 & 0.473&  80.14 &0.000 &  90 &     &SB2O & &$  17.204\pm  0.075 $\\
1: 19B&HD 8956 & 01 25&  60 &7371 & 0.580&  31.98 &0.000 & 100 &     &SB2O &*&$   4.739\pm  0.108 $\\
2: 13A&BD +10 303& 02 13&  14 &4158 & 0.501&   4.58 &0.000 &     &     &SB1O &*&$  -0.911\pm  0.255 $\\
2: 14A&HD 13904& 02 13&  10 &4158 & 0.486&   2.32 &0.000 &  16 & 7443&SB1  & &$  -2.457\pm  1.070 $\\
1: 31B&BD +57 530& 02 13&  26 &3735 & 0.427&   7.26 &0.000 &     &     &SB1O &*&$  11.923\pm  0.084 $\\
2: 15A&BD +28 387s& 02 16&  11 &4466 & 0.460&   2.19 &0.000 &  14 & 7703&SB1  & &$   8.595\pm  0.961 $\\
2: 15B&BD +28 387& 02 16&  30 &4466 & 0.507&  29.24 &0.000 &  36 &     &SB2O &*&$  -6.766\pm  0.323 $\\
2: 16A&HD 14446& 02 17&  32 &4462 & 1.813&  14.29 &0.000 &  38 & 7634&SB2O &*&$  -5.765\pm  1.843 $\\
2: 19B&BD +22 353& 02 26&  22 &4462 & 0.564&   4.61 &0.000 &     &     &SB1O &*&$  19.295\pm  0.100 $\\
2: 20B&BD +17 493p& 03 04&  22 &4469 & 0.997&  43.38 &0.000 &  39 &     &SB2O &*&$ -13.415\pm  0.197 $\\
2: 21B&BD +20 511& 03 06&  32 &4472 & 1.682&   1.64 &0.000 &  37 & 7764&SB1  & &$  25.270\pm  2.718 $\\
2: 24B&HD  23158& 03 40&   5 &1601 & 2.711&   2.38 &0.000 &     &     &VAR? & &$  -0.083\pm  5.770 $\\
2: 33A&HD 27635& 04 21&  26 &4464 & 0.404&   9.83 &0.000 &  34 &     &SB1O &*&$ -39.512\pm  0.188 $\\
2: 33B&BD +63 499& 04 21&  22 &4372 & 0.366&  17.39 &0.000 &     &     &SB1O &*&$ -38.551\pm  0.137 $\\
2: 38B&HD 285970& 04 39&  30 &3735 & 0.390&  51.60 &0.000 &  33 &     &SB2O & &$  13.663\pm  0.172 $\\
2: 40A&HD 33185& 05 06&  23 &4471 & 0.436&   4.95 &0.000 &  31 &     &SB2O &*&$  -3.982\pm  0.214 $\\
2: 41B&HD 241105& 05 06&  16 &3664 & 0.423&  10.98 &0.000 &     &     &SB1O &*&$  16.073\pm  0.068 $\\
2: 48B&HD 59450& 07 27&  20 &7286 & 0.409&   6.47 &0.000 &     &     &SB1O &*&$  -2.379\pm  0.094 $\\
1: 90A&HD  69894& 08 17&   3 & 449 & 0.455&   2.46 &0.003 &     &     &VAR  & &$   0.016\pm  0.912 $\\
1: 93A&HD 71149 & 08 23&  26 &3668 & 0.464&  21.73 &0.000 &  44 & 4087&SB2O &*&$ -11.329\pm  0.093 $\\
2: 54B&HD 80101a& 09 15&  39 &3668 & 0.512&  19.05 &0.000 &  48 &     &SB2O &*&$  55.561\pm  0.172 $\\
1:112A&HD 81997 & 09 26&  52 &6541 & 1.044&   2.35 &0.000 &     &     &SB1O & &$  10.484\pm  0.190 $\\
1:114A&BD +15 2080& 09 33&  32 &3664 & 0.389&   8.74 &0.000 &     &     &SB1O &*&$  -4.562\pm  0.077 $\\
2: 58A&HD 89730& 10 18&  31 &3989 & 1.149&   7.56 &0.000 &     &     &SB1O &*&$  18.837\pm  0.243 $\\
2: 58B&HD 89745& 10 18&  27 &3989 & 1.067&   3.77 &0.000 &  44 & 7283&SB2O &*&$  18.463\pm  0.302 $\\
1:130A&HD  92787& 10 40&  11 &2970 & 3.272&   3.58 &0.000 &  12 &     &SB2  & &$   4.962\pm  5.341 $\\
1:130B&HD 92855& 10 40&  47 &3221 & 0.447&  32.96 &0.000 &     &     &SB1O & &$   4.092\pm  0.087 $\\
1:141A&HD 97815& 11 12&  34 &3669 & 0.324&  36.60 &0.000 &     &     &SB1O &*&$ -11.329\pm  0.065 $\\
2: 62A&HD 98528& 11 17&   4 & 771 & 0.305&   2.94 &0.000 &     &     &VAR  & &$ -11.135\pm  0.777 $\\
2: 64B&HD 100267& 11 29&  34 &4380 & 1.908&  31.48 &0.000 &     &     &SB1O & &$  13.143\pm  0.912 $\\
2: 65B&BD +42 2231& 11 36&  26 &3993 & 0.417&  10.66 &0.000 &     &     &SB1O &*&$  12.415\pm  0.069 $\\
1:156A&HD 102509& 11 45&  12 &5522 & 0.451&  55.54 &0.000 &     &     &SB2O & &$   0.750\pm  0.050 $\\
1:156B&BD +21 2357& 11 45&   9 &3665 & 0.429&   2.51 &0.000 &     &     &SB1  & &$   2.855\pm  1.017 $\\
2: 68B&BD +28 2103& 12 16&  40 &3890 & 0.362&   9.85 &0.000 &     &     &SB1O & &$  17.970\pm  0.065 $\\
2: 70A&HD 109509& 12 32&  17 &5846 & 0.493&   5.72 &0.000 &     &     &SB1O &*&$ -17.949\pm  0.114 $\\
2: 72A&HD 110025& 12 36&  43 &5462 & 0.530&  42.10 &0.000 &     &     &SB1O &*&$  -1.680\pm  0.112 $\\
2: 72B&BD +17 2512& 12 36&  34 &4380 & 0.421&  12.38 &0.000 &     &     &SB1O &*&$ -16.052\pm  0.105 $\\
2: 73B&HD 110106& 12 36&  22 &3884 & 0.366&  10.10 &0.000 &     &     &SB1O &*&$  -8.555\pm  0.087 $\\
2: 74B&BD +26 2401& 12 49&  37 &3894 & 0.483&  26.67 &0.000 &  50 &     &SB2O &*&$  -4.004\pm  0.122 $\\
1:175A&HD 112033& 12 50&  26 &7943 & 0.299&   5.48 &0.000 &  33 &11270&SB1O &*&$  -6.091\pm  0.075 $\\
2: 78A&HD 117044& 13 25&  45 &3909 & 1.603&   2.69 &0.000 &     &     &SB1O &*&$ -12.078\pm  1.735 $\\
2: 79B&HD 117433& 13 26&  49 &4377 & 0.631&  18.69 &0.000 &  79 &     &SB2O &*&$  -9.784\pm  0.164 $\\
2: 81B&HD 234054& 13 35&  23 &3985 & 0.330&   5.53 &0.000 &     &     &SB1  & &$ -43.600\pm  1.784 $\\
2: 83B&BD +37 2460& 13 50&  23 &4255 & 0.391&   7.29 &0.000 &     &     &SB1O &*&$  10.150\pm  0.071 $\\
2: 84B&BD +40 2713& 13 58&  29 &3701 & 0.430&  40.25 &0.000 &     &     &SB1O &*&$ -14.058\pm  0.073 $\\
2: 85B&HIP 69885& 14 16&  18 &3954 & 0.444&   9.42 &0.000 &  20 &     &SB2O &*&$   2.924\pm  0.235 $\\
2: 87A&HD 126661& 14 24&  36 &7002 & 1.425&   2.95 &0.000 &  43 &     &SB2  & &$ -27.832\pm  2.029 $\\
2: 89A&HD 135117& 15 03&  22 &4160 & 0.476&   5.71 &0.000 &  26 & 7305&SB1  & &$   2.026\pm  2.657 $\\
2: 91B&HD 150631& 16 37&  12 &2961 & 2.761&   2.08 &0.000 &     &     &VAR? & &$ -12.958\pm  5.490 $\\
2: 92A&HD 153252& 16 55&  34 &4004 & 0.569&   8.30 &0.000 &     &     &SB1O &*&$ -77.890\pm  0.147 $\\
1:246B&HD 160010& 17 14&  41 &3744 & 0.635&  60.52 &0.000 &  45 &     &SB2O &*&$   6.995\pm  0.115 $\\
2: 94A&HD 158916& 17 28&  35 &4112 & 0.648&   7.06 &0.000 &     &     &SB1O &*&$ -22.512\pm  0.127 $\\
2: 97B&HD 164025& 17 55&  39 &4103 & 0.528&  80.98 &0.000 &     &     &SB1O & &$ -24.130\pm  0.060 $\\
1:258A&HD 167215& 18 11&  22 &7010 & 0.440&   3.15 &0.000 &     &     &SB1O &*&$ -42.805\pm  0.075 $\\
2: 98B&HD 238865& 18 23&  50 &4037 & 0.719&  33.70 &0.000 &     &     &SB1O & &$ -23.459\pm  0.227 $\\
2: 99A&HD 169822& 18 23&  16 &4452 & 0.343&   2.73 &0.000 &     &     &SB1O & &$ -18.977\pm  0.101 $\\
2: 99B&HD 169889& 18 23&  10 &2153 & 0.336&   1.84 &0.000 &  21 & 7406&CST? & &$ -17.952\pm  0.587 $\\
1:280A&HD 194765& 20 24&  41 &2963 & 0.584&  21.04 &0.000 &  72 &     &SB2O &*&$ -15.255\pm  0.081 $\\
1:300B&BD +17 4697p& 22 06&  46 &3764 & 0.643&  62.47 &0.000 &  90 &     &SB2O &*&$  20.963\pm  0.121 $\\
1:307A&HD 214511& 22 34&  52 &3763 & 0.850&  35.57 &0.000 &  80 &     &SB2O & &$  -4.756\pm  0.415 $\\
2:109B&BD +08 4904& 22 35&  43 &4400 & 0.622&  35.32 &0.000 &     &     &SB1O &*&$ -29.873\pm  0.114 $\\
 \hline
\label{tab:Vmoy}
\end{tabular}
\end{minipage}
\end{table*}

Some of the header data are included in Table~\ref{tab:Vmoy}, where the statistical
indicators of variability are also presented.

\section{Calculation of the SB orbits}
\label{sec:SB}

\subsection{The orbits with ``blended'' measurements}

The calculation of the orbital elements of the SB2 systems requires one to answer the preliminary
question: what can we do when the RV of the components are close to the systemic velocity, and when
only one RV is obtained from the CCF since only one dip is visible? 
A solution is to fit the CCF with two Gaussian curves, assuming the widths and the depth ratio of the
CCF components from the observations where they are well separated \citep{Duquennoy87}. However,
the RV then obtained are not very reliable in practice.
Our idea is then to keep the blended measurements as they are, and to use a simple model to express
them as a function of the RV of the components; the orbital elements are then derived from all the measurements.

\begin{figure}
\includegraphics[clip=,height=2.2 in]{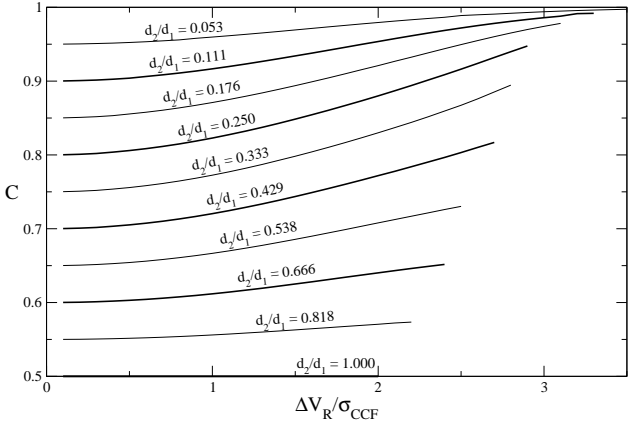}
 \caption{
Simulation of the velocity of a blended measurement with respect to the difference of radial velocity between the components. $C$ is the relative contribution of the primary velocity, $V_1$, to the measured velocity of the blended CCF,
$V_0$. Each line refers to a couple of correlation dips with the same standard deviation, $\sigma_{CCF}$, and with
fixed depth ratio, $d_2/d_1$. The end of the line corresponds to the appearance of two minima in the blended
dip.
$\Delta V_R$ is the absolute value of the difference between the component velocities, $|V_1-V_2|$.
}
\label{fig:C_V2}
\end{figure}

When $C$ is the relative contribution of the primary velocity, $V_1$, to
the measured velocity, $V_0$, we have the relation:

\begin{equation}
V_0 = C\,V_1 + (1-C)\,V_2
\label{eq:V0}
\end{equation}

\noindent
where $V_2$ is the RV of the secondary component.
The radial velocities obtained with {\sc coravel} and with {\sc sophie} are derived by
fitting the CCF with a background level minus a normal distribution \citep{Baranne79}. When the standard deviations and the depths of the CCF of the 
components are fixed, $C$ is a function of 
$\Delta V_R=|V_1 - V_2|$. In order to see the shape of that function, we consider a simple
example hereafter: the same standard
deviation, $\sigma_{CCF}$, is assumed for both correlation dips, and we apply several depth ratios,
$d_2/d_1$. We calculate the sum of two normal distributions and the position of the
blend, $V_0$, is derived by fitting to the sum a single normal distribution. 
The value of $C$ is then derived by inverting
equation~\ref{eq:V0}. These operations are repeated for several separations between
the centers of the two correlation dips, as long as the blended distribution exhibits a sole minimum. 
The results are represented in Fig.~\ref{fig:C_V2}. Each line corresponds to a fixed
depth ratio, which is $d_2/d_1= (1-C(0))/C(0)$, where $C(0)$ is the value of $C$ when the components'
velocities are nearly the same (``nearly'', but not exactly: it is not possible to derive $C$ when the
two dips are perfectly superposed).

\begin{figure*}
\includegraphics[clip=,height=8. in]{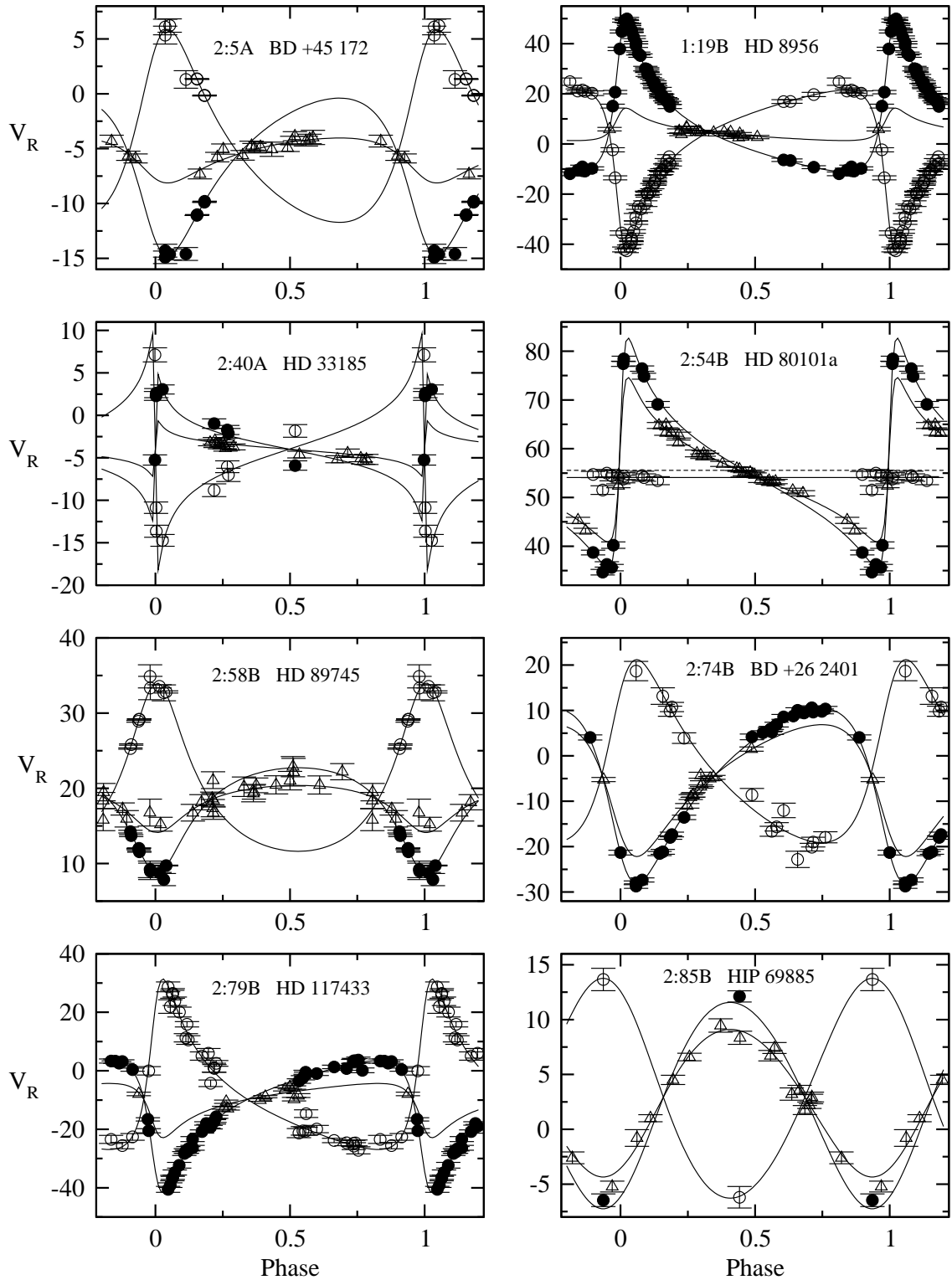}
 \caption{The orbits of the SB2 with at least 10 blended RV measurements, assuming the blends are linear
combinations of the primary and of the secondary velocity.
The black disks are the RV of the primary component, the circles refer to the secondary component, and the
triangles to the measurements obtained from blended CCF peaks. The solid lines refer to the orbital elements.
The system 2:54B is triple, and the secondary is in reality a long period component of a close SB1; the 
RV of the secondary is then fixed; the dashed line is the barycentric velocity of the close binary.
The triple system 1:307A is not 
represented here, since the blended RV are then combination of the primary RV of a short-period SB1, and a
secondary RV with a period much longer; therefore, it is not possible to fold them in phase.
}
\label{fig:orblend}
\end{figure*}

It appears from Fig.~\ref{fig:C_V2} that $C$ could be represented with a polynomial with terms less and less
significant when the order increases: the most important is $C_0$, and a slope may possibly be added, since
$C$ may be as large as $C_0 + 0.1$ when $|V_1-V_2|$ is large. Then, $C=C_0 + C_1|V_1-V_2|$. The $(C_0, C_1)$
coefficients were calculated as terms of the orbital solution for 7 of the 8 systems with more than 10
blended measurements that are presented in Fig.~\ref{fig:orblend} (the triple system 2:54B was set aside). It appeared that $C_1$ was never significant, since its maximum value, 
that was obtained for system 2:74B, was only 2.35 times its uncertainty. Moreover, negative values were found
for 3 systems (1:19B, 2:58B, and 2:79B), although this should not happen in theory. We conclude then that the $C_0$ term is sufficient for deriving
the blended velocities. This is easily confirmed by a visual inspection of Fig.~\ref{fig:orblend}, 
where the model velocity curves derived by assuming only the $C_0$ term in the expression of $C$ are
represented. The 
blended measurements are equally distributed around the theoretical curves, and no deviation related with
$|V_1-V_2|$ is visible in practice.

\subsection{The orbits}

It was possible to derive the orbital elements for 51 stars, including a triple-system solution which
consists of 2 orbits.
The spectroscopic orbits are presented in Tables~\ref{tab:orb1} to \ref{tab:orb3}. We count 40 first
orbits: 27 SB1 and 13 SB2. The 12 other orbits are distributed as follows: Six are new orbits that were
computed taking into account other measurements in addition to ours; these calculations were done
including the offset between the 2 RV sources as a free parameter of the model. Two other orbits 
are published orbits which were
partly based on our measurements; they are just expressed with the same conventions as the others: RV in the
{\sc coravel} system, and epochs in Julian days;
the last 4 orbits are new orbits derived from our measurements alone, since 
including the others did not ameliorate the solution.

The phase plots of 49 orbits are available in electronic form. The two orbits that were
already published are not drawn again.

\begin{table*}
 \centering
 \begin{minipage}{178mm}
  \caption{The orbital elements of the SB up to RA = 9 h 30. The numbers of measurements refer to our measurements, listed in
the RV catalogue presented in Section~\ref{sec:RVcat} plus, where available, 
measurements from external sources.
An asterisk following the CPM name refers to a remark in section~\ref{sec:notes}.}
\scriptsize
  \begin{tabular}{@{}lcccccccccccccc@{}}
  \hline
HD/BD/HIP & $P$ & $T_0$(JD) & $e$ &  $V_0$ & $\omega_1$ & $K_{1,2}$ &  $m_{1,2} \sin^3 i$ or  &  $a_{1,2} \sin i$ &  $N_{1,2}$ & $\sigma(O-C)$ & $C_0$ \\
CPM & (d) & 2400000+ &  & (km s$^{-1}$) & ($^{\rm o}$) & (km s$^{-1}$)  &   $f_1(m)$ (M$_\odot$) &  (Gm) & $N_0$ & (km s$^{-1}$)    \\
  \hline
BD +45 172   & 5556.       &47941.     &0.359 &  -5.484&144.21&  7.18 &1.05     &512.    &  8 & 0.586  & 0.679 \\
2:5A$^*$         &$\pm   79.       $&$\pm   69.     $&$\pm0.069 $&$\pm   0.142$&$\pm  3.11$&$\pm  0.40 $&$\pm0.23     $&$\pm 35.    $& & & $\pm 0.033$        \\
             &             &           &      &        &      &  8.80 &0.86     &628.    &  8 &  \\
             &             &           &      &        &      &$\pm  0.49 $&$\pm0.19     $&$\pm 43.    $&  17 &  \\
&&&&&&&&& \\
HD 4153      &   25.06543  &49976.293  &0.5642& -38.802& 78.45& 19.059&0.01014  &  5.4238& 43 & 0.827         \\
2:7A         &$\pm    0.00118  $&$\pm    0.054  $&$\pm0.0089$&$\pm   0.142$&$\pm  1.38$&$\pm  0.238$&$\pm0.00044  $&$\pm  0.0788$&               \\
&&&&&&&&& \\
HD 5947      &   58.56668  &47871.527  &0.4022&   8.328& 96.21& 10.420&0.00528  &  7.6826&  51 & 0.336         \\
2:8A         &$\pm    0.00505  $&$\pm    0.245  $&$\pm0.0060$&$\pm   0.055$&$\pm  1.27$&$\pm  0.075$&$\pm0.00012  $&$\pm  0.0594$&&               \\
&&&&&&&&& \\
HD 8624      &   14.908342 &49000.156  &0.1342&  17.204& 26.39& 54.375&0.9736   & 11.046 &  44+17 & 0.616 & 0.564 \\
1:18A$^*$    &$\pm    0.000049 $&$\pm    0.025  $&$\pm0.0015$&$\pm   0.075$&$\pm  0.60$&$\pm  0.111$&$\pm0.0047   $&$\pm  0.022 $&& & $\pm 0.035$       \\
             &             &           &      &        &      & 54.573&0.9700   & 11.086 &  43+15 &  \\
             &             &           &      &        &      &$\pm  0.121$&$\pm0.0045   $&$\pm  0.024 $&    3+0 &    \\
&&&&&&&&& \\
HD 8956      &  115.6041   &47976.745  &0.6543&   4.739&316.71& 30.170&0.6308   & 36.267 &  44 & 1.011  & 0.618\\
1:19B        &$\pm    0.0040   $&$\pm    0.105  $&$\pm0.0037$&$\pm   0.108$&$\pm  0.64$&$\pm  0.190$&$\pm0.0100   $&$\pm  0.213 $&& & $\pm 0.018$ \\
             &             &           &      &        &      & 31.756&0.5993   & 38.174 &  40 &          \\
             &             &           &      &        &      &$\pm  0.254$&$\pm0.0082   $&$\pm  0.282 $&  16 & \\
&&&&&&&&& \\
BD +10 303   & 5231.       &47578.     &0.73  &  -0.911&243.  &  4.052&0.0115   &199.2   &  14 & 0.494 \\
2:13A$^*$    &$\pm  747.       $&$\pm   90.     $&$\pm0.33  $&$\pm   0.255$&$\pm 16.  $&$\pm  0.621$&$\pm0.0056   $&$\pm 41.7   $&&               \\
&&&&&&&&& \\
BD +57 530   & 1678.       &46663.     &0.523 &  11.923&243.1 &  4.66 &0.01094  & 91.7   &  26 & 0.297         \\
1:31B        &$\pm   15.       $&$\pm   31.     $&$\pm0.015 $&$\pm   0.084$&$\pm  2.8 $&$\pm  0.10 $&$\pm0.00080  $&$\pm  2.4   $&&               \\
&&&&&&&&& \\
BD +28 387   &   41.2013   &49925.062  &0.3681&  -6.766&  9.87& 23.672&0.2246   & 12.470 &  30 & 1.693         \\
2:15B        &$\pm    0.0033   $&$\pm    0.287  $&$\pm0.0146$&$\pm   0.323$&$\pm  1.18$&$\pm  0.494$&$\pm0.0197   $&$\pm  0.271 $&&               \\
             &             &           &      &        &      & 26.23 &0.2027   & 13.815 &   6 &               \\
             &             &           &      &        &      &$\pm  1.27 $&$\pm0.0154   $&$\pm  0.675 $&     &               \\
&&&&&&&&& \\
HD 14446     &    9.17583  &49982.64   &0.154 &  -5.76 &263.8 & 36.11 &0.0421   &  4.502 &  25 & 7.85  \\
2:16A$^*$     &$\pm    0.00064  $&$\pm    0.58   $&$\pm0.076 $&$\pm   1.84 $&$\pm 24.9 $&$\pm  2.26 $&$\pm0.0083   $&$\pm  0.287 $&&   \\
&&&&&&&&& \\
BD +22 353   &  276.426    &47410.72   &0.277 &  19.295&186.74&  3.744&0.001336 & 13.67  &  22 & 0.432         \\
2:19B        &$\pm    0.362    $&$\pm    6.45   $&$\pm0.035 $&$\pm   0.100$&$\pm  8.66$&$\pm  0.143$&$\pm0.000159 $&$\pm  0.54  $&&               \\
&&&&&&&&& \\
BD +17 493p  &   11.962591 &49995.0938 &0.3965& -13.415&252.67& 64.974&1.1671   &  9.8122&  18 & 1.059 \\
2:20B$^*$    &$\pm    0.000064 $&$\pm    0.0167 $&$\pm0.0035$&$\pm   0.197$&$\pm  0.67$&$\pm  0.331$&$\pm0.0151   $&$\pm  0.0523$&&        \\
             &             &           &      &        &      & 68.330&1.1098   & 10.3189&  17 &        \\
             &             &           &      &        &      &$\pm  0.377$&$\pm0.0142   $&$\pm  0.0594$&     &               \\
&&&&&&&&& \\
HD 27635     &   68.3447   &47817.24   &0.485 & -39.512&180.5 &  5.948&0.000998 &  4.887 &  26 & 0.861\\
2:33A$^*$     &$\pm    0.0159   $&$\pm    0.98   $&$\pm0.035 $&$\pm   0.188$&$\pm  6.9 $&$\pm  0.293$&$\pm0.000162 $&$\pm  0.264 $&&               \\
&&&&&&&&& \\
BD +63 499   & 2630.26     &50113.80   &0.9125& -38.551&242.23& 14.064&0.0520   &208.13  &  22 & 0.508 \\
2:33B        &$\pm    9.97     $&$\pm    1.33   $&$\pm0.0031$&$\pm   0.137$&$\pm  2.67$&$\pm  0.543$&$\pm0.0059   $&$\pm  7.90  $&&               \\
&&&&&&&&& \\
HD 285970    &   56.44579  &49965.668  &0.3627&  13.663&190.88& 26.397&0.6404   & 19.093 &  30 & 0.697         \\
2:38B$^*$    &$\pm    0.00357  $&$\pm    0.217  $&$\pm0.0075$&$\pm   0.172$&$\pm  1.58$&$\pm  0.174$&$\pm0.0248   $&$\pm  0.139 $&&               \\
             &             &           &      &        &      & 35.379&0.4778   & 25.590 &   3 &               \\
             &             &           &      &        &      &$\pm  0.771$&$\pm0.0156   $&$\pm  0.564 $&     &               \\
&&&&&&&&& \\
HD 33185     & 1469.41     &49346.79   &0.9021&  -3.982&271.6 &  8.65 &0.087    & 44.51  &   8 & 1.160 & 0.762\\
2:40A$^*$    &$\pm    0.58     $&$\pm    3.95   $&$\pm0.0090$&$\pm   0.214$&$\pm 17.1 $&$\pm  1.21 $&$\pm0.036    $&$\pm  4.01  $&&  & $\pm 0.061$       \\
             &             &           &      &        &      & 13.95 &0.054    & 60.83  &   8 &         \\
             &             &           &      &        &      &$\pm  1.65 $&$\pm0.023    $&$\pm  4.96  $&  15 &               \\
&&&&&&&&& \\
HD 241105    & 2273.8      &47555.2    &0.4731&  16.073&263.14&  7.939&0.0808   &218.68  &  16 & 0.182         \\
2:41B        &$\pm   10.2      $&$\pm   11.3    $&$\pm0.0214$&$\pm   0.068$&$\pm  2.67$&$\pm  0.235$&$\pm0.0079   $&$\pm  7.15  $&&               \\
&&&&&&&&& \\
HD 59450     & 2708.2      &44826.2    &0.716 &  -2.379&263.19&  4.93 &0.0115   &128.2   &  20 & 0.337         \\
2:48B$^*$=2:49A  &$\pm   32.9      $&$\pm   28.3    $&$\pm0.098 $&$\pm   0.094$&$\pm  5.24$&$\pm  1.07 $&$\pm0.0090   $&$\pm 33.5   $&&               \\
&&&&&&&&& \\
HD 71149     & 1498.3      &48593.4    &0.6876& -11.329&308.84& 13.985&0.8905   &209.20  &  23 & 0.612 & 0.683 \\
1:93A        &$\pm    2.6      $&$\pm    2.2    $&$\pm0.0043$&$\pm   0.093$&$\pm  1.23$&$\pm  0.094$&$\pm0.0305   $&$\pm  2.23  $&&& $\pm 0.032$ \\
             &             &           &      &        &      & 16.320&0.7631   &244.12  &  16 &        \\
             &             &           &      &        &      &$\pm  0.154$&$\pm0.0236   $&$\pm  3.35  $&   5 &               \\
&&&&&&&&& \\
HD 80101a    &   52.81576  &49906.629  &0.7522&  55.561&281.90& 23.53 &0.02041  & 11.262 &  10 & 0.890 & 0.717 \\
2:54B$^*$        &$\pm    0.00077  $&$\pm    0.047  $&$\pm0.0049$&$\pm   0.172$&$\pm  1.27$&$\pm  0.32 $&$\pm0.00077  $&$\pm  0.182 $&& & $\pm 0.023$ \\
&&&&&&&&& \\
HD 81997      & 2815.       &47994.     &0.427 &  10.484&332.7 &  2.79 &0.00469  & 97.7   &  52 & 1.217  \\
1:112A$^*$       &$\pm   46.       $&$\pm   87.     $&$\pm0.092 $&$\pm   0.190$&$\pm 16.8 $&$\pm  0.31 $&$\pm0.00173  $&$\pm 12.1   $&&               \\
 \hline
\label{tab:orb1}
\end{tabular}
\end{minipage}
\end{table*}

\begin{table*}
 \centering
 \begin{minipage}{178mm}
  \caption{Same as Table~\ref{tab:orb1}, for SB with RA between 9 h 30 and 17 h 15.}
\scriptsize
  \begin{tabular}{@{}lccccccccccccccc@{}}
  \hline
HD/BD/HIP & $P$ & $T_0$(JD) & $e$ &  $V_0$ & $\omega_1$ & $K_{1,2}$ &  $m_{1,2} \sin^3 i$ or  &  $a_{1,2} \sin i$ &  $N_{1,2}$ & $\sigma(O-C)$ & $C_0$  \\
CPM & (d) & 2400000+ &  & (km s$^{-1}$) & ($^{\rm o}$) & (km s$^{-1}$)  &   $f_1(m)$ (M$_\odot$) &  (Gm) & $N_0$ & (km s$^{-1}$) &   \\
  \hline
BD +15 2080  &  383.78     &48194.7    &0.1920&  -4.562& 10.31&  5.141&0.00511  & 26.62  &  32 & 0.296         \\
1:114A       &$\pm    0.36     $&$\pm    5.8    $&$\pm0.0145$&$\pm   0.077$&$\pm  5.69$&$\pm  0.100$&$\pm0.00030  $&$\pm  0.52  $&&               \\
&&&&&&&&& \\
HD 89730     &   61.9981   &49934.68   &0.623 &  18.837&201.43& 15.027&0.01045  & 10.02  &  31 & 1.174         \\
2:58A        &$\pm    0.0168   $&$\pm    0.39   $&$\pm0.032 $&$\pm   0.243$&$\pm  3.16$&$\pm  0.858$&$\pm0.00209  $&$\pm  0.66  $&&  &            \\
&&&&&&&&& \\
HD 89745     & 2303.6      &49789.4    &0.3902&  18.463&174.8 &  6.983&0.6876   &203.6   &  10 & 1.081 & 0.788 \\
2:58B$^*$    &$\pm   10.5      $&$\pm   26.8    $&$\pm0.0176$&$\pm   0.302$&$\pm  4.8 $&$\pm  0.183$&$\pm0.0299   $&$\pm  6.3   $&& & $\pm 0.037$ \\
             &             &           &      &        &      & 11.184&0.4293   &326.2   &  10 & &  \\
             &             &           &      &        &      &$\pm  0.197$&$\pm0.0224   $&$\pm  6.6   $&  24 &               \\
&&&&&&&&& \\
HD 92855     &    5.6179028&49997.0514 &0.3203&   4.092&104.62& 21.652&0.005022 &  1.5844&  47+26 & 0.881 \\
1:130B$^*$   &$\pm    0.0000175$&$\pm    0.0114 $&$\pm0.0041$&$\pm   0.087$&$\pm  0.86$&$\pm  0.098$&$\pm0.000070 $&$\pm  0.0074$&& \\
&&&&&&&&& \\
HD 97815     &   86.6079   &49954.45   &0.1966& -11.329&236.4 & 17.172&0.04293  & 20.05  &  34 & 0.339         \\
1:141A       &$\pm    0.0047   $&$\pm    0.33   $&$\pm0.0050$&$\pm   0.065$&$\pm  1.5 $&$\pm  0.089$&$\pm0.00068  $&$\pm  0.11  $&&               \\
&&&&&&&&& \\
BD +12 2343  &    0.786149 &49053.871  &0.023 &  13.14 &296.  & 90.36 &0.0602   &  0.9766&  34 & 5.019      \\
2:64B$^*$    &$\pm    0.000001 $&$\pm    0.076  $&$\pm0.014 $&$\pm   0.91 $&$\pm 35.  $&$\pm  1.32 $&$\pm0.0026   $&$\pm  0.0143$&&               \\
&&&&&&&&& \\
BD +42 2231  &  951.52     &46738.7    &0.0279&  12.415&186.5 &  6.197&0.02349  & 81.06  &  26 & 0.348         \\
2:65B        &$\pm    2.14     $&$\pm   92.7    $&$\pm0.0154$&$\pm   0.069$&$\pm 35.3 $&$\pm  0.099$&$\pm0.00112  $&$\pm  1.30  $&&               \\
&&&&&&&&& \\
HD 102509    &   71.6906   &48431.41   &0     &   0.75 &  0   & 30.12 &0.980    & 29.69  &  12+127 & &\\
1:156A$^*$     &$\pm    0.0004   $&$\pm    0.03   $&  fixed  &$\pm   0.05 $&$\pm  0   $&$\pm  0.07 $&$\pm0.086    $&$\pm  0.07  $&&  &   \\
             &             &           &      &        &      & 33.0  &0.895    & 32.5   & 0+23 &      & \\
             &             &           &      &        &      &$\pm  1.4  $&$\pm0.04     $&$\pm  1.4   $&  &    \\
&&&&&&&&& \\
BD +28 2103  & 2242.3      &50107.0    &0.4402&  17.970&294.25&  4.808&0.01870  &133.11  &40+52& 0.498 \\
2:68B$^*$    &$\pm   18.2      $&$\pm    9.3    $&$\pm0.0118$&$\pm   0.065$&$\pm  2.38$&$\pm  0.070$&$\pm0.00086  $&$\pm  2.19  $&&    \\
&&&&&&&&& \\
HD 109509    & 4091.2      &43192.     &0.0736& -17.949&350.  &  4.019&0.02735  &225.46  &  17 & 0.317         \\
2:70A        &$\pm   51.6      $&$\pm  385.     $&$\pm0.0272$&$\pm   0.114$&$\pm 33.  $&$\pm  0.166$&$\pm0.00342  $&$\pm  9.77  $&&               \\
&&&&&&&&& \\
HD 110025    &   54.87832  &49945.910  &0.3469&  -1.680&184.17& 30.703&0.1361   & 21.730 &  43 & 0.622         \\
2:72A$^*$    &$\pm    0.00140  $&$\pm    0.100  $&$\pm0.0045$&$\pm   0.112$&$\pm  0.80$&$\pm  0.138$&$\pm0.0029   $&$\pm  0.105 $&&               \\
&&&&&&&&& \\
BD +17 2512  &  595.37     &49460.66   &0.3548& -16.052& 58.71&  8.223&0.02810  & 62.94  &  34 & 0.522         \\
2:72B$^*$    &$\pm    1.00     $&$\pm    4.61   $&$\pm0.0158$&$\pm   0.105$&$\pm  3.50$&$\pm  0.168$&$\pm0.00180  $&$\pm  1.35  $&&               \\
&&&&&&&&& \\
HD 110106    & 2899.2      &46789.4    &0.2600&  -8.555&129.90&  6.821&0.0860   &262.59  &  22 & 0.336         \\
2:73B$^*$    &$\pm   20.5      $&$\pm   33.7    $&$\pm0.0172$&$\pm   0.087$&$\pm  3.84$&$\pm  0.152$&$\pm0.0059   $&$\pm  6.26  $&&               \\
&&&&&&&&& \\
BD +26 2401  &   19.43553  &49241.235  &0.3832&  -4.004&130.95& 19.302&0.0499   &  4.764 &  26 & 1.300 & 0.879 \\
2:74B        &$\pm    0.00100  $&$\pm    0.110  $&$\pm0.0100$&$\pm   0.122$&$\pm  1.93$&$\pm  0.193$&$\pm0.0034   $&$\pm  0.042 $&& & $\pm 0.026$ \\
             &             &           &      &        &      & 20.17 &0.0477   &  4.979 &  13 &           \\
             &             &           &      &        &      &$\pm  0.69 $&$\pm0.0018   $&$\pm  0.164 $&  11 &               \\
&&&&&&&&& \\
HD 112033    & 2908.25     &51417.2    &0.6305&  -6.091&347.48&  5.455&0.02287  &169.31  &33 &0.291 &     \\
1:175A$^*$   &$\pm    4.27     $&$\pm    4.4    $&$\pm0.0073$&$\pm   0.075$&$\pm  0.32$&$\pm  0.067$&$\pm0.00109  $&$\pm  3.33  $&&               \\
&&&&&&&&& \\
HD 117044    & 1894.3      &46670.3    &0.669 & -12.08 &  8.51& 20.0  &0.64     &386.    &  45 & 1.76  \\
2:78A        &$\pm   18.4      $&$\pm   30.1    $&$\pm0.149 $&$\pm   1.74 $&$\pm  7.63$&$\pm 14.4  $&$\pm1.43     $&$\pm287.    $&&               \\
&&&&&&&&& \\
HD 117433    &  120.0270   &49437.650  &0.6086&  -9.784&132.70& 22.586&0.486    & 29.58  &  38 & 1.68 & 0.743\\
2:79B        &$\pm    0.0137   $&$\pm    0.226  $&$\pm0.0153$&$\pm   0.164$&$\pm  1.36$&$\pm  0.687$&$\pm0.031    $&$\pm  0.56  $&& & $\pm 0.028$ \\
             &             &           &      &        &      & 29.17 &0.376    & 38.20  &  30 &           \\
             &             &           &      &        &      &$\pm  1.04 $&$\pm0.020    $&$\pm  0.99  $&  11 &               \\
&&&&&&&&& \\
BD +37 2460  & 1235.95     &49480.74   &0.4368&  10.150&306.63&  4.335&0.00761  & 66.27  &  23 & 0.318         \\
2:83B        &$\pm    3.38     $&$\pm    7.15   $&$\pm0.0210$&$\pm   0.071$&$\pm  3.03$&$\pm  0.105$&$\pm0.00061  $&$\pm  1.78  $&&               \\
&&&&&&&&& \\
BD +40 2713  &   13.880363 &47969.8906 &0.2515& -14.058&100.86& 27.735&0.02788  &  5.1236&  29 & 0.370         \\
2:84B        &$\pm    0.000064 $&$\pm    0.0293 $&$\pm0.0038$&$\pm   0.073$&$\pm  0.91$&$\pm  0.119$&$\pm0.00037  $&$\pm  0.0225$&&               \\
&&&&&&&&& \\
HIP 69885    &  912.61     &49402.     &0.092 &   2.924&208.  &  9.41 &0.353    &117.6   &   2 & 0.710 & 0.862 \\
2:85B        &$\pm    4.57     $&$\pm   85.     $&$\pm0.039 $&$\pm   0.235$&$\pm 33.  $&$\pm  0.74 $&$\pm0.100    $&$\pm  9.2   $&& & $\pm 0.041$ \\
             &             &           &      &        &      & 10.02 &0.332    &125.2   &   2 &      \\
             &             &           &      &        &      &$\pm  1.33 $&$\pm0.069    $&$\pm 16.7   $&  16 &               \\
&&&&&&&&& \\
HD 153252    &    5.526174 &47991.656  &0.0302& -77.890&307.3 &  6.547&0.0001609&  0.4973&  34 & 0.681 \\
2:92A$^*$    &$\pm    0.000142 $&$\pm    0.751  $&$\pm0.0320$&$\pm   0.147$&$\pm 48.3 $&$\pm  0.165$&$\pm0.0000121$&$\pm  0.0125$&&               \\
&&&&&&&&& \\
HD 160010    &    5.922546 &49998.328  &0.0295&   6.99 &272.3 & 53.13 &0.934    &  4.325 &  41 & 0.727 \\
1:246B       &$\pm    0.000019 $&$\pm    0.096  $&$\pm0.0029$&$\pm   0.12 $&$\pm  5.9 $&$\pm  0.16 $&$\pm0.017    $&$\pm  0.013 $&&       \\
             &             &           &      &        &      & 82.59 &0.6011   &  6.723 &   4 &         \\
             &             &           &      &        &      &$\pm  0.86 $&$\pm0.0091   $&$\pm  0.070 $&     &               \\

 \hline
\label{tab:orb2}
\end{tabular}
\end{minipage}
\end{table*}

\begin{table*}
 \centering
 \begin{minipage}{178mm}
  \caption{Same as Table~\ref{tab:orb1}, for SB with RA beyond 17 h 15.}
\scriptsize
  \begin{tabular}{@{}lcccccccccccccccc@{}}
  \hline
HD/BD/HIP & $P$ & $T_0$(JD) & $e$ &  $V_0$ & $\omega_1$ & $K_{1,2}$ &  $m_{1,2} \sin^3 i$ or  &  $a_{1,2} \sin i$ &  $N_{1,2}$ & $\sigma(O-C)$ & $C_0$ \\
CPM & (d) & 2400000+ &  & (km s$^{-1}$) & ($^{\rm o}$) & (km s$^{-1}$)  &   $f_1(m)$ (M$_\odot$) &  (Gm) & $N_0$ & (km s$^{-1}$)  &  \\
  \hline
HD 158916    &  941.43     &48155.00   &0.5345& -22.512& 99.84&  6.091&0.01334  & 66.65  &  35 & 0.644         \\
2:94A        &$\pm    2.18     $&$\pm    4.86   $&$\pm0.0221$&$\pm   0.127$&$\pm  3.13$&$\pm  0.151$&$\pm0.00119  $&$\pm  1.99  $&&               \\
&&&&&&&&& \\
HD 164025    &    3.6550085&51089.834  &0.0334& -24.13 &148.8 & 57.64 &0.07256  &  2.895 & 39+55    &  0.53 \\
2:97B$^*$    &$\pm    0.0000015$&$\pm    0.025  $&$\pm0.0014$&$\pm   0.06 $&$\pm  2.5 $&$\pm  0.08 $&$\pm0.00032  $&$\pm  0.004 $&&     \\
&&&&&&&&& \\
HD 167215    & 3518.       &48241.     &0.666 & -42.805&167.2 &  2.64 &0.0028   & 95.    &  22 & 0.337         \\
1:258A       &$\pm   79.       $&$\pm   18.     $&$\pm0.056 $&$\pm   0.075$&$\pm  4.5 $&$\pm  0.32 $&$\pm0.0012   $&$\pm 13.    $&&               \\
&&&&&&&&& \\
HD 238865    &    2.709621 &49997.2337 &0.0171& -23.459&297.  & 34.48 &0.011503 &  1.2845&  29+16 & 1.005  \\
2:98B$^*$    &$\pm    0.000017 $&$\pm    0.1515 $&$\pm0.0065$&$\pm   0.227$&$\pm 20.  $&$\pm  0.21 $&$\pm0.000214 $&$\pm  0.0080$&& \\
&&&&&&&&& \\
HD 169822    &  293.35     &49864.7    &0.586 & -18.977&166.4 &  1.079&0.000020 &  3.52  &16+58& 0.422 \\
2:99A$^*$    &$\pm    0.41     $&$\pm    3.9    $&$\pm0.062 $&$\pm   0.101$&$\pm  8.4 $&$\pm  0.095$&$\pm0.000005 $&$\pm  0.27  $&&    \\
&&&&&&&&& \\
HD 194765    &  160.831    &49251.80   &0.2536& -15.255&106.5 & 16.303&0.3688   & 34.87  &  36 & 0.769 & 0.683 \\
1:280A$^*$   &$\pm    0.057    $&$\pm    0.82   $&$\pm0.0071$&$\pm   0.081$&$\pm  1.8 $&$\pm  0.128$&$\pm0.0086   $&$\pm  0.26  $&& & $\pm 0.024$        \\
             &             &           &      &        &      & 19.297&0.3116   & 41.28  &  31 & &        \\
             &             &           &      &        &      &$\pm  0.211$&$\pm0.0059   $&$\pm  0.43  $&   5 &    &           \\
&&&&&&&&& \\
BD +17 4697p &    9.287228 &50000.650  &0.1812&  20.96 &337.11& 55.60 &0.7236   &  6.982 &  44 & 1.079 &   0.736 \\
1:300B       &$\pm    0.000040 $&$\pm    0.025  $&$\pm0.0028$&$\pm   0.12 $&$\pm  1.04$&$\pm  0.21 $&$\pm0.0068   $&$\pm  0.026 $&& & $\pm 0.061$      \\
             &             &           &      &        &      & 59.60 &0.6750   &  7.484 &  44 & &         \\
             &             &           &      &        &      &$\pm  0.25 $&$\pm0.0058   $&$\pm  0.032 $&   2 &       &        \\
&&&&&&&&& \\
HD 214511AB  &18504.       &49521.     &0.775 &  -4.76 &169.3 &  6.28 &1.624   & 1010.    &  33 & 1.676 & 0.928 \\
1:307A$^*$   &fixed       &$\pm   42.     $&$\pm0.026 $&$\pm   0.42 $&$\pm  3.6 $&$\pm  0.55 $&$\pm0.255    $&$\pm 71.    $&&  & $\pm 0.022$ \\
             &             &           &      &        &      & 11.10 &0.918   & 1786.    &  28 &\\
             &             &           &      &        &      &$\pm  0.55 $&$\pm0.128    $&$\pm122.    $&  19 &   \\
&&&&&&&&& \\
HD 214511A   &    4.570605 &49980.9590 &0.00  &     -  &  -   & 46.49 &0.04756  &  2.921 &  33 & 1.015 \\
1:307A$^*$   &$\pm    0.000023 $&$\pm    0.0069 $& fixed & & &$\pm  0.29 $&$\pm0.00088  $&$\pm  0.018 $&& \\
&&&&&&&&& \\
BD +08 4904  &    7.644897 &47995.0000 &0.0742& -29.873&303.69& 30.909&0.02325  &  3.2403&  43 & 0.732         \\
2:109B       &$\pm    0.000036 $&$\pm    0.0819 $&$\pm0.0055$&$\pm   0.114$&$\pm  3.90$&$\pm  0.164$&$\pm0.00037  $&$\pm  0.0172$&&               \\
\hline
\label{tab:orb3}
\end{tabular}
\end{minipage}
\end{table*}

\subsection{The SB without orbits}
\label{sec:SBnoOrb}

In Table~\ref{tab:Vmoy}, we count 15 stars for which it was not possible to derive a spectroscopic orbit.
In addition, we still found two triple systems (2:16A and 2:98B), with a short-period SB and a drift in the residual RV. 
The figures showing the radial velocities of these 17 stars as functions of the epochs are given in
electronic form. In addition to the two triple systems already mentioned, we still count 7
 long period
SB1 (2:8B, 2:14A, 2:15A, 2:21B, 1:156B, 2:81B and 2:89A), and two long period SB2 (1:130A and 2:87A).
The remaining stars are a pulsating variable (2:5B), 3 stars for which we obtained too few measurements (2:24B,
1:90A, 2:62A), and 2 stars that are probably not variable (2:91B and 2:99B). These objects are discussed in the 
notes in section~\ref{sec:notes} hereafter.

The average velocities of these stars, given in Table~\ref{tab:Vmoy}, are derived from the {\sc coravel} measurements only, since the offset between {\sc coravel} and another system, like {\sc sophie}, cannot be computed. Moreover, for the SB2, only the blended measurements are taken into account. 
The estimation of the uncertainty of the systemic RV is puzzling, since we even do not know when
the range of the RV variations was entirely covered with our measurements. As a consequence,
and unlike the case for a
constant star, the actual error of $\bar {V}$ is not varying as the square root of the number
of the measurements. Therefore, we finally chose to assume as error the standard deviation of 
the RV measurements.

\subsection{Notes on individual objects}
\label{sec:notes}

\noindent
{\bf 2:5A} = BD +45 172. Four {\sc sophie} measurements were taken into account for each component. 
A correction of 3~m~s$^{-1}$ was added to our original estimation of the RV calculated by
fitting two Gaussian curves to the {\sc sophie} CCF. 

\noindent
{\bf 2:5B} = BD +45 171. The star is a semi-regular pulsating variable, and the RV variations are
probably not due to orbital motion.

\noindent
{\bf 2:8B} = BD -01 133. This star was suspected to be a long period SB1 on the basis of the {\sc coravel}
measurements. However, the {\sc sophie} measurements do not confirm this hypothesis, and their variations 
suggest a short period with a small amplitude.

\noindent
{\bf 1:18A} = HD 8624. Revision of the orbit of \citet{Toko99}. The Tokovinin's measurements were taken into
account with a correction of +310~m~s$^{-1}$, corresponding to the
best fit. 

\noindent
{\bf 2:13A} = BD +10 303. Preliminary orbit; our observations cover only 79.5~\% of the period.

\noindent
{\bf 2:14A} = HD 13904. After increasing along the {\sc coravel} observations, the RV is slowly
decreasing over our 6 {\sc sophie} measurements, confirming
a long period binary. Assuming a null offset between {\sc coravel} and {\sc sophie}, one obtains a 
possible period of 8000 days and a periastron around $T=2\,453\,000$.

\noindent
{\bf 2:15A} = BD +28 387s. The {\sc coravel} measurements suggest a long period, and the average
velocity of the {\sc sophie} measurements confirms this hypothesis. A very preliminary orbit was
thus derived, with 
$P =( 8000 \pm 2700)$~days, $T_0=2448000 \pm 800$~JD, $e=0.3 \pm 0.1$, $V_0=( 8.8 \pm
0.4)$~km~s$^{-1}$ and $K_1 =(1.3 \pm 0.3)$~km~s$^{-1}$,
but it does not look reliable, since the period is very uncertain.
However, our 3 {\sc sophie} measurements exhibit variations which suggest that the system
could include a short period component.

\noindent
{\bf 2:16A} = HD 14446. The SB1 orbit was computed discarding all the measurements between 0 and
$-10$~km~s$^{-1}$, which seem to refer to a third component with fixed velocity. The measurements of
the secondary component are not symmetric to those of the primary, and it is impossible to derive a SB2 orbit. 
Finally, a part of a long
period orbit is visible in the large residuals of the SB1 orbit. Therefore, the system could be
quadruple, although the CCF of a sole {\sc sophie} spectrum exhibits one dip only.

\noindent
{\bf 2:20B} = BD +17 493p. Orbit calculated discarding the 4 blended measurements. When they are taken into
account, the blend coefficient is $C_0=0.624 \pm 0.042$, but the orbital elements are not
improved.

\noindent
{\bf 2:21B} = BD +20 511. The {\sc sophie} measurements confirm the variability of the RV. A possible
orbit was found with the following elements: $P=(3070 \pm 125)$~days, $T_0=2446400 \pm 300$~JD, $e=0.5 
\pm 0.3$, $V_0=( 26.7 \pm 0.5)$~km~s$^{-1}$ and $K_1 =(4.8 \pm 1.8)$~km~s$^{-1}$, but it is very uncertain, due to
the large errors of the {\sc coravel} measurements.

\noindent
{\bf 2:24B} = HD 23158. A F5 V type star rather difficult to measure with {\sc coravel}, with very large
uncertainties ($I=2.7$~km~s$^{-1}$). The possible variability is due to the measurement of JD 2448245; when
it is discarded, $P(\chi^2)=12.2$~\%. Therefore, the variability of the star is not certain. If it is
constant, the RV of the star is $\bar{V}=(-1.95 \pm 1.81)$ km~s$^{-1}$.

\noindent
{\bf 2:33A} = HD 27635. The 8 measurements of the secondary seem fixed around -28~km~s$^{-1}$,
and we prefer to discard them. Otherwise, a SB2 orbit is obtained with $K_2=8.3$~km~s$^{-1}$.

\noindent
{\bf 2:38B} = HD 285970. A first orbit was published by \citet{GrifGun}.

\noindent
{\bf 2:40A} = HD 33185. A bright SB2 (6.67 mag) with a semi-major axis expected around 58 mas, which should be
easily separated.

\noindent
{\bf 2:48B} = HD 59450. The star belongs to a triple CPM system and is also 2:49A.

\noindent
{\bf 2:54B} = HD 80101 = ADS 7288AB. A visual binary system with separation 0.3 arcsec. 
The A component is the SB1 with orbital elements in Table~\ref{tab:orb1}. The dip of the
B component is visible on 9 {\sc coravel} CCF, with the fixed velocity
$V_B=(52.08 \pm 0.33)$ km~s~$^{-1}$. Twenty-nine blended RV refer to 
components A and B.

\noindent
{\bf 1:112A} = HD 81997. Revision of the orbit of \citet{DM91}.

\noindent
{\bf 2:58B} = HD 89745. A correction of 0.489 km~s$^{-1}$ was added to the 7 
RV measurements derived from {\sc sophie} for each component, in order to get the best fit. 

\noindent
{\bf 1:130A} = HD 92787. A F5 star with large RV errors. A possible secondary component was
detected on one correlation dip, and it is possible that the other measurements contain blended
observations.

\noindent
{\bf 1:130B} = HD 92855. Revision of the orbit of \citet{Toko94}, which was based on
17 recent measurements, but also on 9 measurements performed between 1916 and
1932. We applied a correction of +0.381 km~s$^{-1}$ for the former, and +1.45 km~s$^{-1}$ 
for the latter.

\noindent
{\bf 2:64B} = BD +12 2343. A first orbit was published by \citet{jeffries95}.

\noindent
{\bf 1:156A} = HD 102509. Orbit of \citet{Griffin04}, partly based on our RV measurements.
The periastron epoch was converted in JD, and the systemic velocity was translated in the {\sc coravel} system.

\noindent
{\bf 1:156B} = BD +21 2357. Drift; the RV was decreasing over 10 years. 

\noindent
{\bf 2:68B} = BD +28 2103. Revision of the orbit of \citet{Latham02}; we found a correction
of -0.120 km~s$^{-1}$ to apply to their measurements.

\noindent
{\bf 2:72A} = HD 110025. A secondary dip was observed by \citet{Halbwachs11}, leading to the mass ratio
$q \approx 0.64$.

\noindent
{\bf 2:72B} = BD +17 2512. A secondary dip was observed by \citet{Halbwachs11}, leading to the mass ratio
$q \approx 0.66$.

\noindent
{\bf 2:73B} = HD 110106. A secondary dip was observed by \citet{Halbwachs11}, leading to the mass ratio
$q \approx 0.75$.

\noindent
{\bf 1:175A} = HD 112033. The star is ADS 8695, a visual binary with $P=359$~yr, $a=1.18$~arcsec and 
$\Delta m= 2.2$~mag \citep{Heintz97}; the secondary component is not visible on our observations, and
the SB1 orbit refers to the brightest component of the visual binary.
A correction of -0.263 km~s~$^{-1}$ was applied to the 7 original
{\sc sophie} measurements.

\noindent
{\bf 2:81B} = HD 234054. A SB1 observed over 11 years, but with a period still longer.

\noindent
{\bf 2:87A} = HD 126661. The RV is slightly decreasing during 6600 days, until the two dips are
separated in our last observations.

\noindent
{\bf 2:89A} = HD 135117. A SB1 observed over 20 years, but with a period still longer.

\noindent
{\bf 2:91B} = HD 150631. The variability status is questionable, since $P(\chi^2)=1.5$~\% when the
measurement of JD 2449931 is discarded. The RV is then ($-12.0 \pm 1.2$)~km~s$^{-1}$.

\noindent
{\bf 2:92A} = HD 153252. A G5-type star without luminosity class. Due to the short period, it cannot
be a giant; assuming the primary component is a dwarf, the secondary component
has a minimum mass around 50 Jupiter masses, and it is a brown dwarf candidate.

\noindent
{\bf 2:97B} = HD 164025. Orbit of \citet{Griffin03}, partly based on our RV measurements.
The periastron epoch was converted in JD, and the systemic velocity was translated in the {\sc coravel} system.

\noindent
{\bf 2:98B} = HD 238865. The star is a triple system, consisting in a long period SB1 with an
additional short period orbit. Preliminary elements of the short period orbit were published 
by \citet{Toko95}; in order to avoid the drift due to the long period, we rejected 21 of our 
{\sc coravel} measurements made before JD 2449000, but we took into account 16 measurements
performed with Russian telescopes; the correction to add to the latter is +1.15 km~s$^{-1}$.

\noindent
{\bf 2:99A} = HD 169822. Revision of the orbit of \citet{Latham02},
with a correction of -0.328 km~s$^{-1}$ to their measurements. The spectral type of the star is G7 V, leading to
a minimum mass around 30 Jupiter masses for the secondary component.

\noindent
{\bf 2:99B} = HD 169889. This star was observed as G141-9 by \citet{Latham02}, who concluded it had a constant RV.
A null value of $P(\chi^2)$ was obtained from all the 10 {\sc coravel} measurements, and also from
the 11 {\sc sophie} measurements. Nevertheless, when one outlying measurement is discarded in both
sets, $P(\chi^2)$ becomes 0.47 and 0.81, respectively. We conclude then that the RV is
probably constant. The RV of the star is then $\bar{V}=(-18.139 \pm 0.114)$ km~s$^{-1}$.

\noindent
{\bf 1:280A} = HD 194765. A bright SB2 (6.70 mag) with a semi-major axis expected around 17 mas, which should be
easily separated.

\noindent
{\bf 1:307A} = HD 214511 = ADS 16111AB. Triple system already studied by \citet{Toko98}. A triple
system solution was computed, combining a long period SB2 with a SB1 as primary component. 
The period of the SB2 was fixed to the value obtained by \citet{Docobo86} for a visual orbit,
as reported by \citet{6orb}.
The assignment of the ``c'' component index in Table~\ref{tab:header} was done as follows: when only
one RV was obtained, it was assumed to be a blend as soon as the difference $|V_1 - V_2|$ was found to be less than 30 km~s$^{-1}$.
The solution presented in Table~\ref{tab:orb3} is based on our measurements only, since it
is better than the one obtained when the measurements of \citet{Toko98} are added.

\section{Conclusion}
\label{sec:conclusion}

An observational program initiated for searching common RV stars among CPM stars resulted in the selection
of 66 stars suspected to be variable. Thanks to observing runs distributed over a very long time
(more than 20 years for 11 stars), we finally derived
a first SB orbit for 40 stars. One of these orbits (2:92A=HD~153252) corresponds to a possible brown
dwarf companion with a minimum mass equal to 50 Jupiter masses. We derive the elements of 13 SB2 orbits,
assuming that the RV of the blended measurements are linear combinations of the RV of the components.

The periods of the 40 new orbits are on average
rather long: more than 1 year for 20 stars, including 5 SB2, and 3 stars (1 SB2) even 
have periods between 10 years and around 15 years. Some of these binaries could be resolved, 
and the masses of the components could be derived from the combination of the visual and the
spectroscopic observations. Among the SB2 brighter than 7~mag, two are ideal targets for
speckle interferometry: 2:40A and 1:280A should have a semi-major axis of
58~mas and 17~mas, respectively. 
Accurate masses should also be
obtained in the future, thanks to the astrometric orbits expected from the forthcoming {\it Gaia} mission
\citep{Halbwachs09}. For that purpose, seven new SB presented in that paper are now measured 
with {\sc sophie} in order to improve their orbits.

In addition to the 40 new SB, we improved the orbital elements of ten others. We also found 11 long-period SB, including 2 SB2, for which it was not possible to derive the orbital elements.

We are now able to select a sample of physical wide binaries, including the components which are themselves close binaries,
and to investigate the statistical properties of these stars and their relations with the other components
of the Galaxy.
This will be the topic of the forthcoming second paper.


\section*{Acknowledgements}

We have benefitted during the entire period of these observations from the support of the Swiss National
Foundation and Geneva Unversity. We are particularly grateful to our technicians Bernard Tartarat, Emile
Ischi and Charles Maire for their dedication to that experiment for more than 20 years.
The {\sc sophie} observations were made thanks to a time allocation of the French {\it Programme National de
Physique Stellaire} (PNPS);
it is a pleasure to thank the OHP staff, and especially Mira V\'eron, for organizing these
observations in service mode.
We are grateful to Andrei Tokovinin for providing an additional RV measurement. We enjoyed to discuss
the method for taking into account the blended measurements of the SB2 with Fr\'ed\'eric Arenou.
Joseph Lanoux and Audrey Morgenthaler made a preliminary calculation of the SB orbital elements.
An anonymous referee made relevant comments, including several corrections of the English.
Rodrigo Ibata did a last reading of the text, still adding a few improvements.
We made use of Simbad, the database of the Centre de Donn\'ees astronomiques de
Strasbourg (CDS).


\label{lastpage}

\end{document}